\documentstyle[preprint,aps,epsf]{revtex}
\tightenlines
\newcommand{\ket}[1]{\left | #1 \right \rangle}
\newcommand{\bra}[1]{\left \langle #1 \right |}

\def\ra{\rangle}
\def\la{\langle}

\def\openone{\leavevmode\hbox{\small1\kern-3.8pt\normalsize1}}
\def\RR{{\rm I\kern-.2emR}}
\def\tr{{\rm tr}\; }
\def\Tr{{\rm Tr}~ }

\def\cb{{\cal B}}

\def\cf{{\cal F}}

\def\ca{{\cal A}}

\def\cc{{\cal C}}

\newcommand{\proj}[1]{\ket{#1}\!\bra{#1}}
\newcommand{\outerp}[2]{\ket{#1}\!\bra{#2}}
\newcommand{\inner}[2]{ \langle #1 | #2 \rangle}
\newcommand{\melement}[2]{ \langle #1 | #2 | #1 \rangle}

\newcommand{\beq}{\begin{equation}}
\newcommand{\eeq}{\end{equation}}
\newcommand{\beqa}{\begin{eqnarray}}
\newcommand{\eeqa}{\end{eqnarray}}
\newcommand{\QED}{\hspace*{\fill}\mbox{\rule[0pt]{1.5ex}{1.5ex}}}

\newtheorem{definition}{Definition}
\newtheorem{theorem}[definition]{Theorem}
\newtheorem{proposition}[definition]{Proposition}
\newtheorem{lemma}[definition]{Lemma}

\newtheorem{possibility}[definition]{Possibility}

\begin{document}

\date{Aug. 31, 2000}
\title{Information-disturbance tradeoff in quantum measurement on the 
uniform ensemble}
\author{Howard Barnum}
\address{Dept. of Computer Science, University of Bristol,
Merchant Venturers Building, Woodland Rd. Bristol BS8 1UB, UK.\footnote{Present address: CCS-3 (Modelling
Algorithms, and Informatics),  
Mail Stop B256, Los Alamos National Laboratories, Los Alamos, NM 87545; 
{\tt barnum@lanl.gov}.}}
\maketitle

\begin{abstract}
I consider the tradeoff between the information
gained about an initially unknown quantum state, and the 
disturbance caused to that state by the measurement
process.  I show that for any distribution of initial states,
the information-disturbance frontier is convex,
and disturbance is nondecreasing with information gain.
I consider the most general model of quantum measurements, 
and all post-measurement dynamics compatible 
with a given measurement.  For the uniform initial distribution 
over states, I show that an optimal information-disturbance
combination may always be achieved by a measurement procedure which
satisfies a generalization
of the projection postulate, the ``square-root dynamics.'' 
I use this to show that the information-disturbance frontier
for the uniform ensemble may be achieved with ``isotropic'' (unitarily
covariant) dynamics.  This results in a significant simplification of
the optimization problem for calculating the tradeoff in this case, 
giving hope for a closed-form solution.  I also show that the discrete
ensembles uniform on the $d(d+1)$ vectors of a certain set of $d+1$ ``mutually
unbiased'' or conjugate bases in $d$ dimensions form spherical
2-designs in $CP_{d-1}$ when $d$ is a power of an odd prime.  This 
implies that many of the results of the paper apply also to these
discrete ensembles. 
\end{abstract}

\vspace{.1in} \noindent
\vspace{.1in}

\section{Introduction}

In this paper, I consider one of the salient ways in
which quantum information differs from classical information
In classical information theory, we may in principle
determine the state of a system arbitrarily accurately with
arbitrarily little disturbance to that state.   By contrast,
in quantum mechanics any measurement which allows 
one to obtain information
about the state of a quantum system must, on average, disturb
that state, except in special cases.
The special cases are  when the possible states of the
system are known in advance to 
lie in one or the other of two or more orthogonal subspaces---then the information
about which of the orthogonal subspaces the state lies in 
can be extracted 
without disturbance.  
This fact underlies some important applications of quantum mechanics
in information processing, notably quantum key distribution 
\cite{Bennett84a}\cite{Bennett92a} and other forms of quantum cryptography, 
as well as some applications to algorithms, such as the proof that PSPACE has
constant-round quantum interactive proof systems \cite{Watrous99a}.
The goal of this 
paper is to quantify the tradeoff between
information gained and disturbance to the system, and derive general
features of that tradeoff.  

In introductory presentations of quantum theory, it is 
often stated that when a quantum system is measured and
a result uniquely associated with a particular eigenvector of the
measured observable is obtained, the system state ``collapses''
to that eigenvector.    This is usually known as ``the 
projection postulate,'' and attributed to 
von Neumann \cite{vonNeumann37a}. 
It clearly represents a disturbance
to the system's state, unless the system is already in an eigenstate
of the measured observable.  A generalization of the projection
postulate to observables with degenerate eigenspaces is known
as ``L\"uders' rule;'' it is slightly different from 
von Neumann's proposed post-measurement dynamics for that 
situation.  L\"uders' rule says that upon a measurement yielding
result $b$ corresponding to a projector $\Pi_b$ (onto a degenerate
eigenspace of the observable) an initial
density operator $\rho$ evolves to 
\begin{eqnarray} \label{eq: Ludersrule}
\hat \rho_{b}^\prime := 
{{\Pi_{b} \rho \Pi_{b} } \over {p_{b}}}\;,
\end{eqnarray} 
where $p_{b} = {\rm Tr} \rho \Pi_{b}$ is the 
probability of obtaining result ${b}$ \cite{Cohen-Tannoudji77a},
\cite{Luders51a}.  Hence the after-measurement 
unconditional density operator becomes $\rho' 
:=  \sum_{b}\Pi_{b} \rho \Pi_{b}$.  But in fact this postulate
describes only one of the many possible ways in which a physical
process of measurement may affect a system.  I will call
measurements in which the effect on the system is described 
by the L\"uders' rule form of the projection postulate
{\em projective measurements}.  Von Neumann's proposal, that the
post-measurement density matrix conditional on observing the 
$b$-th outcome becomes $\Pi_b/\tr \Pi_b$, is another potential 
post-measurement dynamics which is consistent with quantum
theory.  (L\"uders' rule, however, is a more appropriate 
candidate for a ``generalized projection postulate,'' since it 
describes the conditional dynamics of measurement via a projection.)
In the  next section, I
will review
a more  general description both of measurements (as
Positive Operator Valued Measures (POVM's)) and of their 
effects on the system (as a collection of trace-decreasing
completely positive maps, or {\em quantum operations}  summing to
a trace-preserving map).  In this paper, I will generalize
the projection postulate to POVM's.  There are
many collections of operations which are consistent with a
given measurement.    I show
that this generalized ``projection'' postulate selects the 
set of quantum operations which is, on average, least disturbing
to an initially completely unknown input state.  I then 
investigate
the tradeoff between information gained in a measurement, and
expected disturbance of a completely unknown initial state.  
This tradeoff is a quantitative expression of
one of the most salient and distinctive
features of quantum mechanics:  that measurement disturbs a
quantum mechanical system.  

\section{Quantum measurements and operations} 

A  very general
characterization of physically realizable measurement processes
is:  allow the system to be measured to interact unitarily with
another system, often termed the ``ancilla'', which starts out
in some standard state.  Then
measure some
set of orthogonal projectors on the ancilla.  The outcomes of this
measurement may  provide information about the system, and therefore
may be considered to be the results of a measurement on the system.
There is no need to consider the effect of this measurement on the
ancilla if one is only interested in the system, for whether the 
projection postulate, or some other rule, describes what happens
to the ancilla, is not relevant to what happens to the system.
The probabilities of the various results of this measurement, and
the associated change in the system density operator, may be described
solely in terms of the system itself, via the formalism of Positive
Operator Valued Measures (POVM's) and associated ``operations''.

A discrete POVM is  set of positive operators $F_b$ indexed by positive
integers, say, such that
$$\sum_b F_b = I,$$
and the probability of obtaining the measurement result
with index $b$ is ${\rm Tr} \rho F_b$.  For a standard measurement
of a Hermitian observable on the system, the $F_b$ are just the 
projectors onto the eigenspaces of the observable.  Such a 
measurement of projectors is often called ``projection-valued''
(not to be confused with a ``projective'' measurement as defined
above).  I will often call the elements $F_b$ of a POVM ``effects,''
following Ludwig \cite{Ludwig83a} and Kraus \cite{Kraus83a}.  
We will also have some occasion to use continuously indexed POVM's, corresponding
to a continuum of possible measurement results.
These may be loosely thought of as a continuously indexed set of 
``infinitesimal'' positive operators $d\mu(\alpha) F_\alpha$, such 
that $\int d\mu_{\alpha} F_\alpha = I$.  The probability that $\alpha$
lies in a Borel set $\Delta$ is then given by $ \tr \rho \int_{\Delta} 
d\mu(\alpha) F_\{\alpha\}$.

I believe that
confining our attention to discrete, indeed finitely indexed, POVMs and 
instruments results
in no loss of generality.  Arguments similar to, but more involved than,
those of Davies \cite{Davies78a}
and of Ozawa \cite{Ozawa80a} (who treat the maximal information without a disturbance
constraint)
should show that since the optimal information for a given disturbance can 
always be achieved with a POVM having a finite number of outcomes (bounded in 
advance by a polynomial in the dimension of Hilbert space) even when
we initially vary over more general sets of physically reasonable 
POVMs.    Since this promises to be 
rather technical, it will be worked out
elsewhere.
Nevertheless, in Section \ref{sec: isotropic} it will be useful to
use a continuously indexed POVM rather than discrete ones achieving
the same information-disturbance combination, because of the continuous
POVM's greater symmetry.

The general form for the
post-measurement quantum state (density operator) conditional 
on obtaining the result $b$ for a measurement of a POVM consisting
of operators $F_b$ is \cite{Hellwig70a} \cite{Kraus83a} \cite{Choi75a}:
\begin{equation}
\label{eq: operations}
\rho'_b = {\cal A}_b(\rho) = \sum_{i} A_{bi} \rho A_{bi}^{\dagger},
\end{equation}
 where the $A_{bi}$ satisfy 
\begin{eqnarray} \label{eq: consistency}
\sum_i A_{bi}^{\dagger}A_{bi} = F_b\;.
\end{eqnarray}
The linear map $\ca_b$, often referred to as an {\em operation}, 
will be said to have a {\em Hellwig-Kraus (HK) decomposition}, or
simply a decomposition, $\{\ca_{bi}\}$; I will often write this
$\ca_b \sim \{A_{bi}\}_i$.  Note that here and below I use a
convention for ensembles or sets denoted
by expressions within curly brackets.  The convention is that when 
we put part of the expression within the brackets as a subscript
of the right-hand bracket, the overall expression refers to the 
ensemble given by the expression within brackets, when only the 
subscripted piece varies.  Thus for example $\{\rho_{ij}\}$ 
refers to the ensemble of the $\rho_{ij}$ for various $j$ and
fixed $i$.  This is, therefore, the $i$-th in a list of ensembles
indexed by $i$.  (Somewhat irregularly, when there would only be
one subscript and it already appears as the sole subscript of 
the expression within brackets, I will omit it outside the brackets;
thus $\{F_b\}$ means $\{F_b\}_b$.)  I will sometimes 
refer to the operators
of a decomposition as HK operators.

Define $\ca := \sum_b \ca_b,$ (so that 
$\ca(\rho) = \sum_{bi} A_{bi} \rho A_{bi}^\dagger$.
This is  
the overall operation if one does not know the measurement
result; it $\ca$ is trace-preserving.
Notice that $\rho_b^\prime$ is unnormalized, 
and its trace gives the probability of the
measurement outcome.
As usual, I denote a normalized version of an 
operator with a hat:
\beq
\hat{\rho}'_b :=  \rho'_b/\tr \rho'_b\;.
\eeq

I will say that an operation $\ca$ is {\em compatible} with 
a POVM $\{ F_b\}$ if there exists an HK decomposition 
$\{A_{bi}\}$ of $\ca$ such that ($\ref{eq: consistency}$) holds.
The collection of operations $\ca_b$ defined by 
$\ca_b \sim \{A_{bi}\}_i$ is  often referred to as an 
{\em instrument} for the POVM \cite{Davies70a}.
When an operation $\ca$ is viewed as an instrument for a compatible
POVM $\Sigma = \{F_b\}$, I will sometimes call this the {\em procedure}
$(\Sigma, \ca)$;  this is equivalent to the instrument $\{\ca_b\}$.
If we 
use the polar decomposition $A_{bi} =  U_{bi}P_{bi},$
($P$ positive, $U$ unitary), then we have that $F_b = \sum_i 
P_{bi}^2$.  If $P_{bi}$ does not vary with $i$, then all the $P_{bi}$ are 
proportional to ${F_b^{1/2}}$, and with $b$ known 
the value of $i$ contains
no additional information about the initial state.  
If $P_{bi}$ does vary with $i$, then 
the value of $i$ represents further information that is not gathered by the
POVM $\{F_b\},$ but which could have been gathered via a POVM
$\{P_{bi}^2\}$ consistent with the {\em same} operation.  In fact,
one can construct a physical realization of this operation (unitary
evolution on system plus ancilla followed by projective measurement
on the ancilla) such that measuring $F_b$ instead of $P_{bi}^2$
just corresponds to coarse-graining the projective measurement on 
the ancilla by grouping projectors together to form higher-dimensional
ones.  One might expect that
the potential for gathering more information will remove more
quantum coherence, and result in more disturbance of the post-system
state.  The $U_{bi}$ may be thought of as unitary operations that 
the system undergoes conditional on measurement outcomes $b$ and
(if they vary with $i$) on potential measurement outcomes 
$i$ which are not gathered by the POVM  
$\{F_b\}$ but which are nevertheless available to the apparatus, 
so that the further evolution of the state may be conditioned
on them.  If the $U_{bi}$ vary with $i$ while  $P_{bi}$ does not,
then these further ``potential measurement outcomes'' carry
no information about the pre-measurement system state, and
simply represent a stochastic resetting of the state which
is not conditioned on any further information about the 
state---a further noisy disturbance of the state.

A natural generalization of a projective measurement 
is to have a single value
of $i$ in the above sum, and let $A_{bi} = {F_b^{1/2}}$, so that
the unnormalized conditional density operator and the unconditional
post-measurement density operator are given by:
\begin{eqnarray} \label{eq: generalized Luders}
\rho_b^\prime & = & {F_b^{1/2}} \rho {F_b^{1/2}} \nonumber \\
\rho^\prime & = & \sum_b {F_b^{1/2}} \rho {F_b^{1/2}}\;.
\end{eqnarray}
I will say that such measurement procedures exhibit ``the square-root
conditional dynamics,'' and call the associated operation the square-root
operation for that measurement.  Sometimes I will call this 
``the square-root measurement procedure,'' 
although care should be taken not to 
confuse this with the ``pretty-good measurement'', which some authors 
\cite{Eldar2000a}
call the ``square-root measurement''.
In part because of the polar decomposition of the $A$'s just discussed,
we may view any measurement of $\{F_b\}$ as beginning with the 
performance of the square root conditional dynamics, followed, possibly,
by further conditional operations;  this provides one (rather weak) motivation
for thinking of the square root dynamics as the ``minimal disturbance'' one is
compelled to cause.  (It is a weak motivation 
because the subsequent conditional dynamics
can, for some ensembles, be chosen to on average 
repair some of the square root measurement's  damage
to the initial state.)
Even in the 
case of projection-valued measures, the square-root
operation is a very special case, in which the unitaries $U_{bi}$
are all the identity $I$ (up to an irrelevant phase) 
and for each $b$ there is only one $A_b$, which in this case
will just be the projector corresponding to the measurement
outcome.  None of the freedom to add noise by further conditional
unitary operations, or to further disturb the
state by effectively collecting extra information which is
then thrown away, is used in a square-root 
measurement procedure.

\section{Disturbance measures and the information-disturbance 
frontier}

In light of the general formulation of quantum measurement and
its effect on a system, the question arises:  
is there anything special about the projection postulate,
and more generally about the L\"uders
type of measurement?
It is sometimes said, in the context of nondegenerate
Hermitian observables,
that it is the ``least disturbing'' type of 
measurement, since when the measurement is immediately repeated,
one gets the same value of the observable with certainty. 
However, this only means that it doesn't disturb its own eigenstates.
Other states certainly are disturbed, by projection onto the
eigenstates of the observable, and it behooves us to ask whether this
disturbance is in any sense minimal.  If so, one would also like
to know whether ${F_b^{1/2}}$ is the minimal-disturbance generalization
to POVMs. There, it is no longer necessarily
true that repeating the measurement
is guaranteed to give the same result when the operation is 
${F_b^{1/2}}.$ (There is no conditional dynamics which can
provide this guarantee in the case of nonorthogonal $F_b$.)

I will use the fidelity 
$F(\rho,\sigma) := (\tr \sqrt{\rho^{1/2} \sigma \rho^{1/2}})^2$
\cite{Bures69a}, \cite{Uhlmann76a},\cite{Jozsa94b}, 
in specifying a measure
of disturbance for quantum states. 
For pure states $\rho = \proj{\psi}$, this is just 
$\la \psi | \sigma | \psi \ra$.  It is unity when $\rho= \sigma$, 
and zero when their supports are orthogonal.  
It is therefore a reasonable measure of
how similar two quantum states are.
We may define
$1 - F( \rho, \ca(\rho))$ to be the disturbance to the state
$\rho$ by a measurement 
procedure resulting in the operation $\ca = \sum_b \ca_b$:
\beq \label{disturbancedefn}
D :=  1 - F(\rho, \sum_b \ca_b(\rho))\;.
\eeq
Given an ensemble of density operators 
$\{\rho_{\alpha}, \mu(\alpha)\}_\alpha,$ there are several ways one might 
construct a measure of the average disturbance caused by measurement.
For example, one might consider one minus the ensemble average fidelity 
to the input density operator, 
of the post-measurement density operator 
obtained from each ensemble member
by averaging over measurement results:
\beqa
\overline{D}_1 :=
1 - \int d\mu(\alpha) F(\rho_\alpha, \ca(\rho_\alpha))\;.
\eeqa
More reasonable in the context of measurement 
might be consider the fidelity of input density 
operators to the output density operator $\ca_b(\rho_\alpha)$ 
{\em conditional} on the measurement result $b$, averaged 
over both the input ensemble and the measurement result:
\beqa
\overline{D}_2 := 1 - \int d\mu(\alpha) 
\sum_b F(\rho_\alpha ,\ca_b(\rho_\alpha)).
\eeqa
This is disturbance from the point of view of someone
carrying out the measurement, or apprised of its result;
the previous quantity is from the point of view of an outside
observer who does not know the result.
Since $F$ is not linear, these do not define the same quantity;
by the concavity of fidelity \cite{Jozsa94b}, 
$\overline{D}_2 \ge \overline{D}_1$.
One might also consider the disturbance measures
obtained by replacing the
first argument of the fidelity function, 
$\rho_\alpha$ in the above
formulae, by the ensemble average density operator 
$\int d\mu(\alpha) \rho_\alpha.$
These measures seem much less natural (and, again by concavity,
each is less than the
corresponding one of $\overline{D}_1,$ $\overline{D}_2$).
For the case of ensembles of pure input states ($\rho_{\alpha}$ pure),
$\overline{D}_1$ and $\overline{D}_2$ coincide.
For the rest of this paper, I will consider pure input states, and
use this disturbance measure.
This is also the measure used by Fuchs and Peres \cite{Fuchs95b}.  

The average disturbance to 
an initial pure state, where the average is taken over 
some ensemble of pure states specified by a probability measure
$\mu(\ket{\psi}),$ on Hilbert space, is given by   
\begin{eqnarray}
\overline{D}  :=    
1 - \int d\mu(\ket{\psi})
\sum_b F( {\cal A}_b(|\psi\rangle \langle \psi|),
|\psi \rangle \langle \psi |) \nonumber \\ 
 =  1 - 
\int d\mu(\psi) 
\sum_{bi} |\langle \psi | A_{bi} |\psi \rangle|^2\;.
\end{eqnarray}
The ensemble I will be most concerned with is
$d\mu(\ket{\psi}) = d\Omega_\psi,$ the unitarily invariant measure on Hilbert
space, normalized to integrate to $1$.

To measure the information gained about an initial ensemble 
$\Psi \sim d\mu(\ket{\psi})$ , I will use the mutual 
information between the prior distribution and the measurement
outcome, denoted $H(\Psi:B)$.  Note that $\Psi$ is a random variable
taking Hilbert space vectors as values, and distributed according to 
$d\mu(\ket{\psi})$;  $B$ is a random variable taking 
measurement results $b$ as values, distributed according to  
$p(b|\ket{\psi}) = \tr F_b \proj{\psi} = \melement{\psi}{F_b},$ conditional
on the initial state $\ket{\psi}$.  
The information gain is:
\begin{equation}
H(B:\Psi) =  H(B) - H(B|\Psi).
\end{equation}
The second term is the average, using the measure $d\mu(\ket{\psi})$
over states $\ket{\psi},$ of the conditional information
\beqa
H(B|\ket{\psi}) := -\sum_b p(b|\ket{\psi}) \log{p(b|\ket{\psi}}\;.
\eeqa

I will also occasionally consider a different measure of disturbance,
involving the {\em entanglement fidelity} 
\begin{eqnarray}
F_e(\rho,{\cal A}) :=  \sum_{bi} |\mbox{tr} A_{bi} \rho|^2\;.
\end{eqnarray}
The entanglement fidelity of a density operator $\rho$ under an 
operation $\ca$ is less than or equal to the average pure-state
fidelity of any ensemble for $\rho$ under $\ca$ \cite{Schumacher96a}.
I will define the
entanglement disturbance $D_e(\rho, \ca)$ to be $1 - F_e(\rho, \ca).$  
It is an upper bound to the pure-state disturbance to any ensemble for
the density operator $\rho$.
Since it is defined for
an initial density operator rather than an initial ensemble, it
is less suitable than (\ref{disturbancedefn})
for use in a information-disturbance relation 
like that described above, since the information gain against 
which disturbance is graphed involves a particular initial ensemble.
However, it does provide a lower bound to the information-disturbance
frontier.
(We could of course just fix some ensemble, such as the ``Scrooge''
ensemble for the density operator $\rho$ \cite{Jozsa94c},
which is the one
about which the minimum information is gained, and plot information
gain against minimum entanglement disturbance for 
this particular fidelity
measure.  One might speculate that the entanglement disturbance would provide a
reasonably tight bound on the disturbance to the Scrooge ensemble.)

Given a measurement and a known initial state $|\psi_0\rangle,$ 
it is easy to 
come up with an operation, consistent with the measurement,
which minimizes the pure-state 
disturbance (\ref{disturbancedefn}):
just set the state back to its initial value no matter what.
This may be accomplished by letting $A_{bi} = \lambda_{bi}
|\psi_0\rangle \langle  b i\rangle,$ where $\lambda_{bi}$
and $| b i \rangle$ are the eigenvalues and eigenvectors
of ${F_b^{1/2}}.$ (It is easily checked that this measurement
has average fidelity one, and satisfies 
the criterion (\ref{eq: consistency}) for compatibility with
the POVM $\{F_b\}$.)
But this measurement will severely disturb other initial states.
When we set up our measuring apparatus we may or may not know
anything about the states we are going to be measuring.  
A fair way of assessing whether an operation corresponding to a
set of effects is minimally disturbing, without assuming any
prior knowledge about the state to be measured, is to minimize the 
disturbance averaged over initial pure states with the unitarily
invariant measure.  This also makes the problem of finding the 
least disturbing measurement analytically 
tractable.

Ultimately, one would like to find the
{\em information-disturbance frontier} for a given ensemble, 
defined as the graph of minimal disturbance 
for a given
amount of information collected about the initial state, 
against information collected.  (We could equivalently define
it via the dual optimization problem, 
as the graph of maximal information collectable by a measurement
causing no more than a fixed amount of disturbance, against that
disturbance.) Formally, we must define this graph as the infimum
of disturbance for a given amount of information collected about
the state, and show that this infimum is in fact attainable.

Short of an explicit 
expression (which seems unlikely for a general ensemble), one 
would like to derive general properties of 
this frontier---such as the fact that 
minimal disturbance increases with 
information collected.  
This may appear obvious: one could
argue that 
we couldn't cause less disturbance by collecting more information,
for then one could just collect the smaller amount of information 
by doing an experiment that would collect more information with
less disturbance, but
adding noise to the readout, or not looking at all details of the
answer.  Fuchs and Peres 
\cite{Fuchs95b} have explored this frontier for two-state
ensembles, with possible applications to quantum cryptography.

Some progress toward the structure of the information-disturbance
frontier may be made by noting that both disturbance
measures considered
above (and indeed also all the disturbance measures which 
are one minus an average ensemble fidelity), are linear in the
operation, and the information is linear in the POVM.  More 
precisely, from a set of POVM's $\{F^i_b\}$ (where $i$ 
indexes which POVM and $b$ indexes which operator in the 
POVM) and  
associated sets of tracepreserving operations $\{ {\cal A}^i\}$
indexed by $i$
with 
operator decompositions $\{A^i_b\}$
we construct the POVM's and operations 
which are convex combinations of these:
\beqa
\{ G_{ib} \} :=  \{ \lambda^i F^i_b\}, \\
{\cal B} \sim \{ \sqrt{\lambda^i} A^i_b \}.
\eeqa
Then for any of the disturbance measures discussed above 
($1 - F_e(\rho, \ca)$ and $1 - \overline{F}(E,\ca)$, regardless
of the density operator $\rho$ or the ensemble $E$ used in 
the average), we have
\beqa
D({\cal B}) = \sum_i \lambda^i D({\cal A}^i)\;.
\eeqa
Also, for {\em any} ensemble of states:
\beqa
\overline{H}(\{ G_{ib} \} ) = \sum_i \lambda^i \overline{H}(\{ F^i_b \} ).
\eeqa  
where the overbar indicates the ensemble average over the information
conditional on the input state.  Hence, given any two points in 
the information-disturbance feasible set, the line joining them is
entirely within the set.  This implies 
\begin{theorem}
The information-disturbance
frontier $D(I)$ for a pure-state ensemble is convex.  
\end{theorem}
(Our convention is that
a function $f$ is {\em convex} if $\lambda f(x) + (1 - \lambda)f(y)
\ge f(\lambda x + (1-\lambda) y)$, 
i.e. the average of the
function is greater than or equal to  the function of the average.)

Since the disturbance measures
under consideration are positive, and one endpoint of $D(I)$ is
at the origin, this implies 
that the information-disturbance frontier for a pure-state ensemble
is nondecreasing:
minimal disturbance
is nondecreasing with information obtained.  
That is, 
\begin{proposition} \label{shape}
For any disturbance measure of the form
$1 - \overline{F}(E,\ca)$
or 
$1 - F_e(\rho, \ca)$
the minimal disturbance required to obtain a given amount
of information about some fixed ensemble (which need not even
be that used in the disturbance measure) is nondecreasing in 
the amount of information obtained.
In fact, it may
have a flat section following the zero-information endpoint, 
but at some point must become and remain monotonically increasing.
\end{proposition}

We may use this fact to show:
\begin{proposition} 
For any pure-state ensemble,
if
a point on the upward-sloping portion of the information-disturbance
frontier is attainable then it is attainable
by a POVM $\{F_b\}$ measured in such a way
that the conditional operations $\{\ca_b\}$ may be taken to have 
a one-operator decomposition.
\end{proposition}

We will say such a measurement procedure has {\em one-term conditional 
dynamics}.
  
{\em Proof:}
Consider an operation $\ca$ compatible with  
a POVM $\Sigma$, and suppose that the procedure $(\Sigma  , \ca)$ achieves
some point $\zeta$ on the upward-sloping portion of 
the information-disturbance frontier.  
Let $(\Sigma, \ca)$ have multiterm conditional dynamics. 
Then $\ca$ 
is also compatible with some
POVM $\Upsilon$ which finegrains $\Sigma$, such that $(\Upsilon,
\ca)$ exhibits one-term conditional dynamics.  $\Upsilon$
gathers no less 
information than $\Sigma$.
If it gathers the same amount, then $\zeta$ is achievable by 
$(\Upsilon, \ca)$ and the proposition is established for the point
$\zeta$.  If it gathers more information than $\Sigma$, then by the
strict monotonicity of this portion of the frontier (Proposition 
\ref{shape}), $\ca$ must have disturbance greater than the 
minimal disturbance for $\Sigma$, contradicting the assumption that
it was least-disturbing for $\Sigma$, and so establishing the proposition.
\QED

So in investigating measurement procedures achieving the 
information-disturbance frontier, we may confine
our attention to those with a single $A_b$ for each POVM element
$F_b$.

In fact, we can also show that for any feasible 
information-disturbance
combination $(D,I)$, there exist ways of achieving $(D,I')$ and
$(D',I)$ with one-term conditional dynamics, where $D' < D$ and $I'> I$.
The first is done by considering the fine-grained POVM of the proof
above;  the second by mixing this with the trivial POVM, $I,$ measured
with one-term conditional dynamics.  
This enables us to confine our attention, when considering the
form of the information-disturbance frontier, to measurement procedures 
exhibiting one-term conditional dynamics, even without any assumption that
the frontier is attainable.  

However, one might also wish
to directly show the superiority of the single-term operations
for {\em arbitrary} 
POVM's, 
and possibly even for ensembles other than the uniform one.
That is, one might hope to show 
\begin{possibility} \label{one-term always best}
For any POVM and any 
pure state 
ensemble, the set of 
operations least-disturbing to that ensemble and compatible with
that POVM contains an operation with one-term conditional 
dynamics. 
\end{possibility}
One might even try to show that the least-disturbing operations
compatible with a POVM {\em all} have one-term conditional 
dynamics.  (To show this, the definition of one-term conditional 
dynamics would have to modified so as to include, at least,
``trivial'' multiterm conditional dynamics in which the many 
Kraus operators $A_{bi}$ have, when polar decomposed, the same
isometric part, and positive parts proportional to each other.)
Multiple-term operations consistent with the same POVM involve
potentially collecting more information, and so it seems reasonable
that this would cause more disturbance.  Conceivably, however, it might
cause less disturbance if the 
additional information helped restore
the initial state better than could be done without
it.
 
It appears difficult to establish the desired property in general, 
but we may show it for the uniform ensemble.   
(It is easy to show if our disturbance
measure, instead of an ensemble average fidelity,  is one minus the
entanglement fidelity of the uniform density operator;  this
is done in Appendix \ref{appendix: min entanglement disturbance}.)

\begin{theorem} \label{oop bop sh'bam}
One-term conditional dynamics always give a
minimally-disturbing way of measuring a given POVM, on the 
uniform ensemble. 
\end{theorem}

Consider the contribution
to $\overline{F}$ from a particular value of $b$:
\begin{eqnarray} \label{eq: partialaveragefid}
\sum_i \int d\Omega_\psi |\langle \psi| A_{bi} |\psi \rangle|^2 \equiv
\int d\Omega_\psi |\langle \psi| {\cal A}_b(|\psi \rangle \langle 
\psi|) |\psi \rangle\;.
\end{eqnarray}
The disturbance in the multi-term case separates into terms for each
$A_{bi} \equiv U_{bi} P_{bi}$ in which $i$ indexes the different 
operators corresponding to 
the outcome $b$, $U_{bi}$ is unitary and $P_{bi}$ positive (the polar
decomposition again).
From this and the result of \cite{vonNeumann37a}
(cf. \cite{Horn85a}) that for $X \ge 0,$ 
$\max_{\mbox{\scriptsize unitary } V} |\tr VX|$ occurs 
where $VX = \sqrt{X^\dagger X}$, it follows that
$| A_{bi} |\psi \rangle|^2$
is maximized where $U_{bi} = I$, so $A_{bi} = P_{bi}^{1/2}$.

We therefore proceed by a proposition which will be
proved below.
\begin{proposition} \label{oh frabjous day}
For any $|\psi\rangle$ and positive $P_1, P_2$
\begin{equation}
\label{eq: conjecture}
\langle \psi | P_1 | \psi \rangle^2
+ \langle \psi | P_2 | \psi \rangle^2 \le
\langle \psi | \sqrt{P_1^2 + P_2^2} | \psi \rangle^2\;.
\end{equation}
\end{proposition}
This implies that $A_b=\sqrt{F_b}$ is
a minimally disturbing operation to $\Omega$
for general POVM's, since any 
(finite) purportedly better set of operations can be repeatedly 
coarse-grained in the 
manner of Equation {\ref{eq: conjecture}} to arrive at $\sqrt{F_b}
\equiv \sqrt{\sum_{bi} P_{bi}^2}.$  This proves Theorem 
\ref{oop bop sh'bam}.  \QED 

In fact, Proposition (\ref{oh frabjous day}),
implies that for $any$ initial ensemble, not just the uniform one,
coarse-graining
the measurement decreases the disturbance caused by a measuring
with square-root conditional dynamics.  
However, this does not yet prove that
coarse-graining a measurement decreases the minimal disturbance for
an arbitrary ensemble, for the minimally disturbing operation compatible
with a given POVM  will generally not be the square-root operation
unless the ensemble is uniform.  

For our application, we also have $\sqrt{P_1^2 + P_2^2} \le I$, but
the proposition holds more generally.
Proposition \ref{oh frabjous day} is 
not hard to prove when the $P_{bi}$ commute.
Let $P_1$ have (positive) eigenvalues $\lambda_i$.  Let $P_2$ have (positive) eigenvalues $\eta_i$ for the same
eigenvectors as $P_1$, so that they commute.
Then
$\sqrt{P_1^2 + P_2^2}$ commutes with them, and has 
positive eigenvalues
$\sqrt{\lambda_i^2 + \eta_i^2}$ and the same eigenvectors.
We will use these eigenvectors as a basis and write the 
inequality in components, with $x_i$ being
the $i$-th component of $|\psi\rangle$ in this basis.
The desired inequality (\ref{eq: conjecture}) becomes:
\beqa
(\sum_i   x_i^2 \lambda_i )^2 + (\sum_i x_i^2 \eta_i )^2
\le
(\sum_i x_i^2 \sqrt{\lambda_i^2 + \eta_i^2})^2 \\
\sum_{ij} x_i^2 x_j^2 \lambda_i \lambda_j + \sum_{ij}
 x_i^2 x_j^2 \eta_i \eta_j \nonumber \\
\le
\sum_{ij} x_i^2 x_j^2 \sqrt{(\lambda_i^2 + \eta_i^2)
(\lambda_j^2 + \eta_j^2)} \\
\sum_{ij} x_i^2 x_j^2 (\lambda_i \lambda_j +
\eta_i \eta_j) \nonumber \\
\le 
\sum_{ij} x_i^2 x_j^2 \sqrt{\lambda_i^2\lambda_j^2 + 
\eta_i^2 \eta_j^2 + \lambda_i^2 \eta_j^2
 + \lambda_j^2 \eta_i^2}.
\eeqa
Rewriting the LHS as 
\beqa
\sum_{ij} x_i^2 x_j^2 \sqrt{(\eta_i \eta_j + \lambda_i \lambda_j)^2}
\nonumber \\
= \sum_{ij} x_i^2 x_j^2 \sqrt{\lambda_i^2 \lambda_j^2 + \eta_i^2 \eta_j^2
+ 2 \lambda_i \lambda_j  \eta_i \eta_j}.
\eeqa
we see that if 
$$
\lambda_i^2 \eta_j^2
 + \lambda_j^2 \eta_i^2 \ge
 2 \lambda_i \lambda_j  \eta_i \eta_j,
$$
then the LHS is less than the RHS.  And this is indeed
the case:  letting 
$a= \lambda_i \eta_j$ and $b = \lambda_j \eta_i$,  
it reduces to the fact that $a^2 + b^2 \ge 2 a b$
(which is true since $(a-b)^2 \ge 0$, with equality 
iff $a=b$).  Equality in our expression occurs when 
$\lambda_i \eta_j = \lambda_j \eta_i$ for all $i, j,$
that is, when $\lambda_i/\lambda_j = \eta_i/\eta_j$.
In other words, the POVM elements $P_{bi}^2$
are proportional to 
each other.  This implies that knowing which of them occured gives
us no additional information about the state.

{\em Proof of Proposition \ref{oh frabjous day}:}
In the general case
Proposition \ref{oh frabjous day} 
follows quickly from the following theorem of
T. Ando \cite{Ando79a}, which is easily seen to be equivalent to Lieb's
concavity theorem (\cite{Lieb73b}; see also discussions
in \cite{Bhatia97a}, especially p. 273, and\cite{Ahlswede99a}).
\begin{theorem}[Ando]
For $0 \le t \le 1$, the map:
\beq
(A,B) \rightarrow A^t \otimes B^{1-t}
\eeq
is jointly concave on pairs of positive operators $A,B$.
\end{theorem}
{\em Proof of Proposition \ref{oh frabjous day}:}
Consider the map from operators to the reals given by:
\beqa
\cf(A) = \melement{\psi}{A^{1/2}}^2\;.
\eeqa
Then (\ref{eq: conjecture}) is equivalent to the superadditivity of
$\cf$:  $\cf(A) + \cf(B) = \cf(A + B)$ on the cone of positive operators
(let $A = P_1^2, B = P_2^2$).
Since $\cf$ is linearly homogeneous ($\cf(\lambda A) = \lambda \cf(A),$)
this is equivalent to the concavity of $\cf$ on the unit
interval.
Also, $\cf(A) \equiv \bra{\psi}\bra{\psi} A^{1/2} \otimes A^{1/2}
\ket{\psi}\ket{\psi}$.  Ando's theorem has as a special case
the concavity of the mapping $A \rightarrow A^{1/2} \otimes A^{1/2},$
which implies that any diagonal matrix element of it (in any
basis) including that between $\ket{\psi}\ket{\psi}$ and itself, 
is a concave function.  (Ando's theorem holds on the entire cone of 
positive operators,
which is why we did not need
the restriction $\sqrt{P_1^2 + P_2^2} \le I$ in Proposition 
\ref{oh frabjous day}.) \QED

\section{Minimally-disturbing operations compatible with a given measurement}
\label{mindisturbance}

With arbitrary POVMs, with the operation for
each measurement outcome given by a single decomposition
operator $A_b$, we can
show that $A_b = {F_b^{1/2}}$ is a minimal-disturbance operation and
evaluate the minimal disturbance.
That is,
\begin{theorem} \label{maintheorem}
Let $\{F_b\}$ be a POVM, and let
$\{\ca_b\}$ be a set of operations compatible with that $POVM$.
If each $\ca_b$ has an operator decomposition consisting of a
single operator, then 
\beqa
\sum_b &\int& d\Omega_{\psi} F(|\psi\ra\la\psi|, \ca_b(|\psi\ra\la\psi|))
\nonumber \\
&\le& \sum_b \int d\Omega_{\psi} F(|\psi\ra\la\psi|, 
{F_b^{1/2}}|\psi\ra\la\psi| {F_b^{1/2}}) \nonumber \\
&=&\frac{1}{d(d+1)}(d + \sum_b (\tr {F_b^{1/2}})^2 \;.
\eeqa
\end{theorem}

The proof proceeds via the following Lemma, which also appears
with a different proof in \cite{Ashikmin2000b}.

\begin{lemma} \label{integrationlemma}

Define 
\beq
\Pi :=
\int d \Omega_{\psi} \outerp{\psi}{\psi} \otimes \outerp{\psi}{\psi}
\eeq
Then 
\beq
\Pi
= \frac{1}{d(d+1)}
\sum_{ij} \outerp{i}{i} \otimes \outerp{j}{j} + \outerp{i}{j} \otimes 
\outerp{j}{i}\;.
\eeq
\end{lemma}

\noindent
{\em Proof of Lemma:} 
\beqa
\label{baluebebop}
\int d\Omega_{\psi} \outerp{\psi}{\psi} \otimes \outerp{\psi}{\psi}
\nonumber \\
= \int d\Omega_{\psi} 
\sum_{ijlm} \inner{i}{\psi} \inner{\psi}{j} \inner{l}{\psi} 
\inner{\psi}{m} \outerp{i}{j} \otimes \outerp{l}{m} \;.
\eeqa
With the notation $\inner{i}{\psi} = r_i e^{i \phi_i}$, etc..., 
the $ijlm$-th matrix element of $\Pi$ may be written as:
\beq
\int d \mbox{\boldmath $r$} d \mbox{\boldmath $\phi$} 
\delta( | \mbox{\boldmath $r$}| - 1) r_i e^{i \phi_i}
r_j e^{-i \phi_j}
r_l e^{i \phi_l}  r_m e^{-i \phi_m}\;.\\
\eeq
Here $d \mbox{\boldmath $r$} = dr_1 dr_2 \cdots dr_d$, 
$d\mbox{\boldmath $\phi$} = d \phi_1 \cdots d \phi_d.$

The angular integrals give zero except in three cases, for which 
the matrix elements in (\ref{baluebebop}) are as follows:
\beq
\begin{array}{ll}
1.~i=j, l=m, i \ne l: &   \int d\Omega_{\psi} 
|\inner{i}{\psi}|^2 |\inner{l}{\psi}|^2 \\
2.~i=m, j=l, i \ne j: & \int d\Omega_{\psi} 
|\inner{i}{\psi}|^2 |\inner{j}{\psi}|^2 \\
3.~i=j=l=m\;: & \int d\Omega_{\psi}|\inner{i}{\psi}|^4\;.
\end{array}
\eeq
The integrals are easily done using Eq. 
(12)
of Jones \hbox{\cite{Jones91a}}, which yields:
\beqa
\int d\Omega_{\psi} |\inner{\psi}{a}|^2|\inner{\psi}{b}|^2
= \frac{1 + |\inner{a}{b}|^2}{d(d+1)}\;,
\eeqa
where $\ket{a},\ket{b}$ are {\em any} normalized, but not 
necessarily orthogonal or identical, vectors. 
For our cases $1$ and $2,$ the matrix elements are
$1/d(d+1)$;  case $3$ gives $2/d(d+1)$. 
We combine $1/2$ times the case $3$ terms 
with each of case $1$ 
and $2,$ enabling us to remove the inequality
condition on the indices, and change the dummy index $l$ to 
$j$ to obtain the Lemma. \QED

\noindent
{\em Proof of Theorem \ref{maintheorem}}:

Note that 
\beqa \label{eq: projectorintegrates}
\int d\Omega_{\psi} 
\melement{\psi}{A} \melement{\psi}{B} \equiv 
\tr (\Pi (A \otimes B))~,
\eeqa
Lemma \ref{integrationlemma} enables one to write this as
$(1/d(d+1))\sum_{ij} \left(\melement{i}{A} \melement{j}{B} + 
\bra{i}A\ket{j} \bra{j}B\ket{i}\right)$.
Hence the average overlap becomes:
\beqa
\frac{1}{d(d+1)}
\sum_b \sum_{ij} \left(\melement{i}{A_b} \melement{j}{A_b^\dagger} + 
\bra{i}A_b \ket{j} \bra{j}A_b^\dagger\ket{i} \right) \nonumber \\
= \frac{1}{d(d+1)} \sum_b 
\left(|\tr A_b|^2 + \tr A_b A_b^\dagger)\right)\;.
\eeqa
By the linearity and cyclicity of the trace and
the fact that $\ca$ is 
trace-preserving ($\sum_b A_b^\dagger A_b = I$), the second term
in parentheses is $d$.  We wish to maximize this overlap
(thereby minimizing disturbance) over all single-term operations 
compatible with $F_b$.  So for each $b$, we maximize the $b$-th 
term 
over all $A_b$ such that $A_b^\dagger A_b = F_b$.  By the 
polar decomposition of operators, such $A_b$ have the form
$U_b F_b^{1/2}$.  From this and the result of \cite{vonNeumann37a}
(cf. \cite{Horn85a}) that for $A \ge 0,$ 
$\max_{\mbox{\scriptsize unitary } V} |\tr VA|$ occurs 
where $VA = \sqrt{A^\dagger A}$, it follows that 
$|\tr A_b|$ is maximized where $U = I$, so $A_b = F_b^{1/2}$.  Thus
the optimum overlap is obtained with the square root conditional
dynamics.
It is given by:

\beqa
\overline{F}_{max} = \frac{1}{d(d+1)} (d + \sum_b (\tr F_b^{1/2})^2)\;.
\eeqa
The corresponding minimal disturbance is 
\beq
\label{eq: minimum disturbance for arbitrary POM}
\overline{D}_{min}
= 1 - \overline{F}_{max}\;. 
\eeq
\QED

Consider the special case of 
effects proportional to one
dimensional projectors.
The effects $F_{b}$ become $g_b |b\rangle \langle b|$, 
where $g_b$ are proportionality constants satisfying $\sum_b g_b = d$.
The optimum overlap and disturbance for the uniform ensemble,
with one-term conditional dynamics, are given by:
\begin{equation}
\label{eq: distmin}
\overline{F}_{max} = {2 \over {d+1}},~
\overline{D}_{min} = {{d-1} \over {d+1}}\;.
\end{equation}

\section{Information} \label{sec: information}

We have found, in Eq. 
(\ref{eq: minimum disturbance for arbitrary POM}), the minimum   
disturbance for measurement of an arbitrary POVM.  This is a
step towards deriving the information-disturbance frontier.
As a special case, we found the minimal disturbance
to be $(d-1)/(d+1)$
for a class of measurements in which the effects 
are proportional to one-dimensional projectors.
At the opposite pole from these ``fine-grained measurements'' is the
ultimate coarse-grained measurement of a single effect which is the 
identity operator.  This yields zero information, and can be accomplished with no disturbance. These extreme cases presumably represent the endpoints
of the information-disturbance frontier.
Another step toward deriving the frontier is to find the information
gained in measurements of the fine-grained type investigated above,
which is clearly greater than zero, as is the disturbance they cause.
This will pin down the maximal-information endpoint.  It 
turns out that the information yield is the same for all
such fine-grained measurements, whether the effects are orthogonal
or not.  This is a special case of the fact that any fine-grained 
measurement gives the same information about the ``Scrooge'' ensemble.
(The Scrooge ensemble for a given density operator $\rho$
is defined as the ensemble (from among ensembles for $\rho$) 
for which the accessible information is minimal \cite{Jozsa94c}. 
The uniform ensemble
is the Scrooge ensemble for the uniform density operator $I/d$.)  

Here I present a different derivation
of the information gained by a finegrained measurement,
which applies to the the uniform ensemble only and 
uses the methods of Jones \cite{Jones91a}.
Recall that the information gain from measurement is the mutual
information between the prior distribution and the measurement
outcome, denoted $H(\Psi:B)$.  I will use this in the form:
\begin{equation}
H(B:\Psi) =  H(B) - H(B|\Psi).
\end{equation}
This can be calculaated form 
the prior probability measure on states 
$p(|\psi\rangle)$ which we assume to be 
the unitarily invariant one, 
and the conditional probabilities
$p(b|\psi)$ of the data (measurement outcomes) given the initial
state, which are ${\rm Tr} g_b |b\rangle \langle b | \psi \rangle 
\langle \psi | = g_b |\langle b | \psi \rangle | ^2 .$  (Here I 
use the notation for finegrained measurements
introduced at the end of Section \ref{mindisturbance}.) 
The first term is 
\begin{equation}
H(B) = - \sum_b p(b) \log{p(b)}.
\end{equation}
Since 
\begin{equation}
p(b)  \equiv  \int d\Omega_{\psi} p(|\psi\rangle) p(b|\psi) \\
 =  \int d\Omega_{\psi} g_b  |\langle b | \psi \rangle | ^2 \\
 =  {g_b \over d},
\end{equation}
\
\begin{equation}
H(B)  =  - \sum_b {g_b \over d} \log {g_b \over d} \\
 =  
\label{eq: unconditinfo}
-{1 \over d} \sum_b g_b \log{g_b} + \log{d}, 
\end{equation}
where I have used equation (7) of \cite{Jones91a} to do the 
integral, and have also made use of the fact that $\sum_b g_b = d.$

The second term is:
\begin{eqnarray}
H(B|\Psi) & = & - \int d\Omega_{\psi} \sum_b p(b|\psi) 
\log{p(b|\psi)} \\
& = & - \int d\Omega_{\psi} \sum_b  g_b  |\langle b | \psi \rangle | ^2
\log{ g_b  |\langle b | \psi \rangle | ^2} \\
& = & - \sum_b g_b \int d\Omega_{\psi} |\langle b | \psi \rangle | ^2
(\log{g_b} + \log{ |\langle b | \psi \rangle | ^2}) \\
& = & -\sum_b g_b \log{g_b}  
\int d\Omega_{\psi} |\langle b | \psi \rangle | ^2 \nonumber \\
& - & \sum_b g_b  \int d\Omega_{\psi} |\langle b | \psi \rangle | ^2 
\log{ |\langle b | \psi \rangle | ^2}\;. \nonumber \\
{} 
\end{eqnarray}
The first integral is the same one we encountered in $H(B)$, and its
value is $1/d.$  The second integral is more complicated, but can be 
done using the same formula as the first (or see \cite{Wootters90a});
its value is  
\beq
-{1 \over d} \sum_{k=1}^{d-1} { 1 \over 1 + k }.
\eeq
Hence
\begin{equation}
\label{eq: conditinfo}
H(B|\Psi) = {1 \over d} \sum_b g_b \log{g_b} 
+\sum_{k=1}^{d-1} { 1 \over 1 + k }.
\end{equation}
Combining equations (\ref{eq: conditinfo}) and (\ref{eq: unconditinfo}),
we obtain
\begin{equation}
H(B:\Psi) =  \log{d} - \sum_{k=1}^{d-1} { 1 \over 1 + k }.
\end{equation}

This depends only on $d$, and not on the weights $g_b$;  as
long as the $F_b$ are proportional to one-dimensional projectors,
the information gained about a maximally uncertain initial pure
state is the same, whether the measurement is of orthogonal projectors
or some other set of maximally fine-grained effects.  

Unfortunately, finding the information gain from measuring an 
arbitrary POVM is a much more difficult problem.

\section{The information-disturbance frontier} \label{sec: frontier}
For the information-disturbance frontier, we need the information 
gain maximized over possible measurements and compatible operations
causing a given
level of disturbance (or less).  Equivalently, we need the minimal disturbance 
measurement and associated operation which gives a fixed level of 
information gain.  Since the minimal disturbance 
associated
with all fine-grained measurements is the same, and they all yield the
same information gain, we have found the high-information 
endpoint of the information-disturbance frontier. For any other 
set of effects will be a blurring
(by allowing positive operators not proportional to projectors)
or coarsening (by allowing higher-dimensional projectors) of these
effects, resulting in less information gain and the possibility
of less disturbance.  Clearly, the other endpoint is at zero information
and zero disturbance, achieved by the identity operation of doing nothing.
One might speculate that the minimally disturbing
measurement (for the uniform ensemble) for any given level of 
information obtained, is to measure a fine-grained set 
of effects with some probability, and otherwise to do nothing.
That is, our POVM is given by the set $\{\alpha I, (1-\alpha)
F_b\},$ where the $F_b$ form a fine-grained POVM.
Then the tradeoff frontier is a straight line between the known
endpoints.
However, it seems unlikely that the frontier is perfectly straight.
This would just be too boring to be true.  In the next section, we
will make some progress towards obtaining a closed form for the 
information-disturbance frontier, by showing that for each 
point on the frontier, there exists an optimal measurement
procedure associated with a very simple operation, that of swapping
in the maximally mixed state with some probability and otherwise
leaving the state undisturbed.  (This operation is not compatible
with the measurement just discussed, that gives the straight-line frontier.)

\section{Towards the full frontier} \label{sec: isotropic}

We will say an operation $\ca$ is {\em unitarily covariant} if
\beq
W^\dagger \ca(W \rho W^\dagger) W = \ca(\rho)\
\eeq
for any unitary $W$.
We will also introduce a convention for ensembles or sets denoted
by expressions within curly brackets.  The convention is that when 
we put part of the expression within the brackets as a subscript
of the right-hand bracket, the overall expression refers to the 
ensemble given by the expression within brackets, when only the 
subscripted piece varies.  Thus for example $\{\rho_{ij}\}$ 
refers to the ensemble of the $\rho_{ij}$ for various $j$ and
fixed $i$.  (This is, therefore, the $i$-th in a list of ensembles
indexed by $i$.)

Using the unitary invariance of the ensemble $\Omega$, 
we will show that

\begin{theorem} \label{theorem: unitarily covariant optimal measurement}
There is always a {\em unitarily covariant}
way of obtaining a given $I$ with minimal disturbance to $\Omega$.

In other words, for this measurement and
conditional dynamics the operation $\ca := \sum_b \ca_b$
is unitarily covariant.

\end{theorem}

{\em Proof:}
By the unitary invariance of the ensemble $\Omega$, for any 
fixed unitary $U$, the POVM 
$\{UF_bU^\dagger\}$ has the same information and the same 
minimal disturbance
(for $\Omega$) as $\{ F_b\}$.  (This is so because
the information depends only on the
probabilities $p_b = \melement{\psi} {U F_b U^\dagger}$, so
transforming the POVM is equivalent to transforming the ensemble,
which we know is invariant. Similarly, the minimally 
disturbing conditional dynamics compatible with this POVM
are given by the operation with decomposition $A_b = 
(UF_bU^\dagger)^{1/2} \equiv U F_b^{1/2}U^\dagger.$ The average 
disturbance depends on the $A_b$ only through 
$\melement{\psi}{A_b}
\equiv \melement{\psi} {U F_b^{1/2} U^\dagger}$, so again we may view
the unitary transformation as applied to the ensemble, which is 
invariant under it.)
By the linearity of disturbance and information in the POVM and
operation, respectively, the continuously indexed POVM 
\beq
\{ d\mu(U) UF_b U^{\dagger} \}_{b,U}
\eeq
where POVM elements are indexed by {\em both} $b$ and $U$, 
achieves the same information and disturbance as $\{ E_b \}$.
Here $d\mu(U)$ is the (unitarily invariant) Haar measure on the
unitary group $U(d)$.
This POVM is unitarily invariant in the sense that applying any
unitary $V$ to all elements of the POVM just results in the same
POVM with the elements reindexed.  
The optimal associated operation is given by the continuous
decomposition:
\beq
\{ d\mu(U)^{1/2} UF_b^{1/2}U^\dagger \}_{b,U}\;.
\eeq
The ``square root of a measure'' here is just formal notation.
(For a rigorous treatment of such operations 
as ``Radon-Nikodym derivatives of quantum instruments'', see
\cite{Davies76a}, \cite{Holevo98a}.)
This operation is defined by its action:
\beqa \label{rassamahdoody}
\ca(\rho) = \sum_b \int {d\mu(U)} U F_b^{1/2} U^\dagger 
\rho U F_b^{1/2} U^\dagger\;
\eeqa
note that the formal square root does not appear here.
The unitary covariance of $\ca$ 
is straightforward from (\ref{rassamahdoody}) and the 
unitary invariance of $d\mu$. 
\QED

A unitarily covariant (which we will also call {\em isotropic})
operation may be viewed as 
mixing in the uniform density 
operator with some probability $p$ (cf. e.g. \cite{Werner98a},
\cite{Zanardi98a},\cite{Keyl99a}):

\beq
\ca_p(\rho) = (1-p) \rho + p (I/d)\;.
\eeq
This operation causes disturbance
\beq
D_{min}(\Omega, \ca_p) = p \frac{d-1}{d}\;.
\eeq

To calculate the information-disturbance frontier, we now
need only to calculate the maximum, over POVMs compatible with
the isotropic operation 
$\ca_p$, of the information gathered by the POVM.  Here I will
not give a closed form for the maximum, but I will give an 
approach which reduces the problem from a constrained maximization
to an unconstrained one.
To do this, we recall some more of the basic theory of 
quantum operations.
By saying a POVM is ``compatible with the operation'' $\ca$, 
we mean that it can be
measured by an instrument which gives rise to that operation
(when measurement results are averaged over).
Any POVM compatible with operation $\ca$ is given
by coarsegraining some 
set of operators $\{F_{b}\}$ defined by $F_b = A_b^\dagger A_b$
for some decomposition $\{A_b\}$ of the operation $\ca$.
We need only consider the POVM's obtained as $\{F_b := A_b^\dagger A_b\},$
and not the coarsegrainings, since the coarsegrainings obtain less (or at
least no more) information.  Thus every decomposition of an operation determines
a compatible POVM, and all compatible POVM's are obtained by this procedure
(plus coarsegraining).  

Any two decompositions of the same operation,   
$\{A_i\}$ having $r$ operators
and $\{B_i\}$ having $s$ operators, 
are related by \cite{Choi75a}:
\beqa \label{eq: unitary remixing}
A_i = \sum_{j=1}^s m_{ij} B_j
\eeqa
where $m$ is the matrix of a maximal partial isometry from the complex
vector space $\cc^s$ to 
$\cc^r$.
A partial isometry is a generalization of a unitary 
operator, which must satisfy $V V^\dagger = \Pi$ for some projector
$\Pi$.  Such an isometry will then also satisfy $V^\dagger V = \Gamma$
for some projector $\Gamma$ having the same rank as $\Pi.$
If the range and domain spaces
of a linear operator $V$ 
have different dimensions, it will not be possible to
find a unitary mapping between the two:  the best one can do is find a
partial isometry $V$ such that one of $V  V^\dagger$ and $V^\dagger V$ is 
the identity (whichever one operates on the smaller space).    
We 
will call such a map a {\em maximal} partial isometry between
the spaces $S_1$ and $S_2$.  
A partial isometry with $ V V^\dagger$ (and  hence $V$)
having rank
$C$ may be thought of as projecting onto a $C$-dimensional 
subspace of $V$'s domain
Hilbert space and then mapping that subspace
unitarily to a $C$-dimensional subspace of the range Hilbert
space.  
Thus if $s \le r$ in (\ref{eq: unitary remixing}), 
$m$'s columns are 
$s$ orthonormal vectors in $\cc^r$:
\beqa
\sum_j m_{ij}^* m_{kj} = \delta_{ik}\;
\eeqa
or in other words:
\beqa
m m^\dagger  = I^{(s)}\;.
\eeqa

Any quantum operation on a system $Q$ 
may be realized \cite{Stinespring55a},\cite{Hellwig70a},\cite{Kraus83a} 
by a ``unitary representation''
in which the Hilbert space $Q$ is extended by adjoining an environment
$E$ prepared in a standard state $|
0^E\rangle$, and the system and environment undergo a unitary interaction,
followed by a projection on the environment system.  Any such unitary
interaction with a given initial environment state 
determines a quantum operation.  (In the case of a 
trace-preserving operation, the environment projection is the identity.)
That is,
\beqa 
\ca(\rho) = \tr_E (\pi^{E} U^{QE}|0^E\ra\la 0^E| \otimes \rho^{Q} 
U^{\dagger QE} \pi^{E})\;.
\eeqa
The operators $A_i$ in the operator decomposition representation discussed
above, turn out to be the ``operator matrix elements''
\beqa
A^Q_i = \la i^E| U^{QE} |0^E\ra
\eeqa
of the unitary interaction, between the initial 
environment state and orthonormal
environment vectors 
$|i\ra$ of the basis used for the partial trace over the environment.
The freedom (\ref{eq: unitary remixing}) 
to ``unitarily mix'' the operators $A_i$, 
obtaining another valid decomposition, 
is just the freedom to do the enviroment partial trace in a 
different environment basis (related to the first by the transpose
of the unitary used in remixing).
See \cite{Schumacher96a} for a more extended discussion of this.
Here, we merely emphasize that in order to get {\em all} decompositions
as we vary the measurement on the environment, it was assumed that
the environment was initially in a pure state.

The import of this for our problem of extracting information about $\ket{\psi}$
via measurements compatible with $\ca_p$ is that 
we may vary over the relevant ``finegrained'' POVMs compatible
with $\ca_p$ by  
imagining we implement $\ca_p$ with an initially pure environment, and 
varying over {\em all} measurements on the environment.
We may do this by letting the interaction 
$U^{QE}$ swap
half of bipartite a maximally entangled
state from the environment into the system $Q$, conditional on 
``quantum dice'' loaded with probability $p$.
\footnote{``Quantum dice'' are usually taken to 
consist of a pure entangled state of two systems,
used as dice by conditioning operations on some third system on
the eigenbasis of one of the two entangled systems.  The resulting 
operation on the third system has the effect of randomly performing
one of the operations which were performed conditionally, with 
probabilities given by the eigenvalues of the reduced density matrix
of the entangled state.  Below we use a slightly different formulation 
which applies to our special case of either doing or not doing some
operation.  This involves an extra ``flag'' dimension of the environment
instead of an extra environment qubit. 
It reduces the required number of Hilbert space dimensions, because we 
don't have to have the maximally entangled state ready for partial swapping
even in that subspace where the swapping won't be done, as we would if we
conditioned on a qubit value.}
Since half (i.e., 
one subsystem) of 
a bipartite maximally entangled state has the uniform density operator
$I/d$, this just replaces the state of $Q$ with the uniform density 
operator, with probability $p$.  In other words, it effects the 
isotropic operation with parameter $p$.
In more detail, we let the environment be the $(d^2 +1)$-dimensional
Hilbert space
\beq
E = E_1 \otimes E_2 \oplus F\;,
\eeq
where
$E_1 \cong E_2 \cong Q$ are $d$-dimensional and $F$ is 
a one-dimensional ``flag''
on which the swapping is conditioned. 
We prepare an initial environment state
\beq \nonumber
|0^E\ra = \sqrt{1-p} \ket{F} + \sqrt{p} \sum_{i=1}^d
\frac{1}{\sqrt{d}}\ket{i^{E_1}} \ket{i^{E_2}}
\eeq
and realize the operation $\ca_p$ on $Q$ through the unitary
interaction:
\beq
U^{QE}:= (SWAP(E_1, Q) \otimes I^{E_2}) \oplus (I^F \otimes I^Q)\;.
\eeq
SWAP simply swaps the states of $E_1$ and $Q$; it is defined by:
\beqa
SWAP(E_1,Q) \ket{j^Q}\ket{i^{E_1}}= \ket{i^Q}\ket{j^{E_1}}\;, 
\eeqa
so that, overall
\beqa
U^{QE}\ket{j^Q}\ket{i^{E_1}}\ket{k^{E_2}} = \ket{i^Q}\ket{j^{E_1}}
\ket{k^{E_2}} \\
U^{QE}\ket{j^Q}\ket{F^E} = \ket{j^Q}\ket{F^E}\;.
\eeqa
When $|\psi^Q\rangle$ goes in on the measured system, the final 
{\em environment} state is 
\beqa
\rho^{E'}(|\psi\rangle, p) = (1-p) \proj{F^{}} + p \proj{\psi^{E_1}} \otimes
\frac{I^{E_2}}{d} \nonumber \\
+ \sqrt{\frac{(1-p)p}{d}}\left(\ket{F} \bra{\psi^{E_1}} \bra{\psi^{E_2}}
+\ket{\psi^{E_1}} \ket{\psi^{E_2}} \bra{F} \right)
\eeqa
Now, any information about the {\em initial} state of $Q$ obtainable
by a measurement compatible with $\ca_p$ may be obtained by measuring
the environment $E$ after the above-defined interaction $U^{QE}$, for
each such measurement made on the environment after the interaction 
corresponds, via the unitary representation of operations,  to a 
decompositions 
$\{A_b\}$ of the operation $\ca_p,$ and thus to a POVM on Q compatible
with $\ca_p$, and as we vary over all measurements on an initially
pure $E$ we obtain fine-grainings of all such POVM's.

The uniform distribution $\Omega$ for initial states 
$\psi$ gives rise, via the dynamical evolution $U^{QE}$, to a distribution
$\mu_p$
on final environment states $\rho^{E'}$.  
The accessible information about $\rho^{E'}$ 
is the maximal information 
obtainable about $|\psi\rangle$ by measurements on $E'$
consistent with this 
operation, and hence gives us the maximal information about 
the initial preparation $\ket{\psi}$ 
consistent with
the isotropic operation $\ca_p$.  As we vary $p$ 
parametrically, we get the information-disturbance frontier
for the uniform pure-state ensemble $\Omega$.   

\section{Spherical 2-designs}
All the results of this paper (notably, Theorems \ref{oop bop sh'bam}
and \ref{maintheorem})
which involve only average pure-state fidelities over the uniform ensemble 
(and not, for instance,  information), hold also for a class
of discrete pure-state ensembles.  These ensembles are the 
{\em spherical t-designs} for $t \ge 2$ 
in $d-1$-dimensional complex projective space
$CP_{d-1}$ (that is, the space of rays of the 
$d$-dimensional Hilbert space, isomorphic
to the space of pure quantum states $\proj{\psi}$).  
Various equivalent definitions of these
designs exist, but the one relevant here is that a spherical $t$-design 
is a finite set $\Delta \subset CP_{d-1}$ 
such that the uniform integral over $CP_{d-1}$
of a polynomial $P$ of degree
no higher than $t$ is equal to the discrete average of the polynomial 
evaluated on the points of the design:
\beq
\int_{CP_{d-1}} P(\pi) = \frac{1}{|\Delta|} \sum_{\pi \in \Delta} P(\pi)\;.
\eeq
(As usual, $|S|$ denotes the cardinality of a set $S$.)
A reasonably good supply of small (size quadratic in the dimension)
spherical $2$-designs exists, and some are
given by the following construction.
Define two orthonormal bases to be {\em unbiased} \cite{Wootters89a} 
or {\em conjugate} \cite{Wiesner83a} if 
any inner product of a vector from one basis with one from the other
has modulus $1/\sqrt{d}$.
There exist sets of ``complementary''
bases which are higher-dimensional analogues of the eigenbases of 
$\sigma_x$, $\sigma_y$, and $\sigma_z$.  These are the ``mutually unbiased
bases''  (MUBs) introduced by  
Ivanovi\'{c} \cite{Ivanovic81a}
(for prime dimension), and by Wootters and Fields \cite{Wootters89a} 
(for prime power dimension).
Let 
the index 
$k=0,...N-1$ specify which basis; $i=1,...,d$ specifies
which vector in the basis.  A set of $N$ orthonormal bases 
indexed by $k$ is
said to be {\em mutually unbiased} \cite{Wootters89a} or {\em conjugate}
\cite{Wiesner83a} if for all $k \ne l$
\beqa
|\inner{e^k_i}{e^l_j}| =  1/\sqrt{d}\;.
\eeqa

For $d=p^n$, $p$ prime, Wootters and Fields constructed
$d+1$ mutually unbiased 
bases $\ket{e^k_i}.$  The construction uses the finite field $F_{p^n}$ of 
prime power order, also known as Galois fields $GF(p^n)$, which has 
$p^n$ elements (including zero).
For odd primes, the construction is as 
follows.
One basis may be chosen arbitrarily;  in this ``standard'' basis
the $l$-th component of the $j$-th vector of the $k$-th basis is:
\beqa
\inner{l}{e^k_j} = \frac{1}{\sqrt{d}}\omega^{{\rm Tr~}[kl^2 + jl]}
\eeqa
where $l,k,j$ range over the $p^n$ elements of $F_{p^n}$.
\beq
\omega := e^{2 \pi i /p}\;,
\eeq
(a primitive $p$-th root of unity)
and
\beq
{\rm Tr~}[x] := x + x^p + x^{p^2} + \cdots + x^{p^{n-1}}\;.
\eeq
Note that the trace has values in a subfield of $F_{p^n}$
isomorphic to $F_p$.
Verifying that these are mutually unbiased is a relatively
calculation using elementary properties of the trace 
on finite fields \cite{MacWilliams77a}, \cite{Lidl86a})
and
Gauss sums \cite{Lidl86a}.
In particular, the properties ${\rm Tr~}(x+y) = 
{\rm Tr~}(x) + {\rm Tr~}(y), (x,y \in F_{p^n})$
and ${\rm Tr~}(cx) = c {\rm Tr~}(x), c \in F_p, x \in F_{p^n}$
are fundamental.
Wootters and Fields also give a construction for $p=2$,
but it is more complicated and I will not present it here. 
Working independently of Wootters and Fields and of Ivanovi\'{c},
and using ideas from coding theory and finite geometry,
Calderbank, Cameron, Kantor, and Seidel \cite{Calderbank97a}
also found sets of
$d(d+1)$ 
mutually unbiased bases for prime-power dimension, which may well
be the same as Wootters' and Fields'.  (At least some cases were
also found by other authors cited in \cite{Calderbank97a}.)
Calderbank et. al. also state that many unitarily inequivalent
such sets of MUBs must exist.  (Constructions are known at least
for $d$ a power of 2.)
$d+1$ meets an upper
bound (valid for arbitrary $d$) on the number of such bases, 
established by Delsarte, Goethals, and Seidel \cite{Delsarte75a}.
I know of no examples meeting the bound for $d$ with 
distinct prime factors.  

\begin{theorem}
The set of $d(d+1)$ vectors $\ket{e_i^k}$ 
belonging to the union of the 
$(d+1)$ mutually unbiased (aka conjugate) bases
constructed by Wootters and Fields for $d$ a power of an odd prime, is a 
spherical $2$-design in $CP_{d-1}$.
\end{theorem}

\noindent
{\em Proof:}

Any second-degree polynomial in $\pi = \proj{\psi}$ may be written 
$\sum_\alpha \tr \pi A_\alpha \tr \pi B_\alpha$ 
for some finite set of linear operators $A_\alpha$ and $B_\alpha$.
(This is shown e.g. in  \cite{Rains97a}, or using Lemma 1 from
\cite{Grassl98a}.) 
So by (\ref{eq: projectorintegrates}) we need only show that:
\beqa \label{upsilondefined}
\Upsilon := \frac{1}{d(d+1)}\sum_{ki} \proj{e_i^k} \otimes \proj{e_i^k} = \Pi\;.
\eeqa
First consider the operator made by summing 
over all basis vectors of all the MUBs except
the standard basis:
$\Lambda := \sum_{ki \in F_{p^n}} \proj{e_i^k} \otimes \proj{e_i^k}$.
In the standard basis this has matrix elements 
\beqa
& &\bra{\alpha}\bra{\gamma} \Lambda 
\ket{\beta}\ket{\delta} \nonumber \\
&=& \sum_{ki} \inner{\alpha}{e^k_i}
\inner{e^k_i}{\beta}\inner{\gamma}{e^k_i}\inner{e^k_i}{\delta}
\nonumber \\
&=&(1/d^2) \sum_{ki} \omega^{
\Tr(k \alpha^2 + i \alpha) -
\Tr(k \beta^2 + i \beta) +
\Tr(k \gamma^2 + i\gamma) -
\Tr(k \alpha^2 + i \delta) 
}
\nonumber \\
\label{scooby dooby doo wah day}
&=& (1/d^2) \sum_k \omega^{\Tr k(\alpha^2 - \beta^2 + \gamma^2 - \delta^2)}
\sum_i \omega^{\Tr i(\alpha - \beta + \gamma - \delta)} 
\;.
\eeqa
We thus have a product of two sums of the form $\sum_{k \in F_{p^n}} \omega^{kx}$.
This sum is easily shown to be equal to $p^n \delta_{x,0}$.
(By definition $\delta_{x,0}=0$ if $x=0$, $1$ otherwise.)
To show it, note \cite{MacWilliams77a} that as 
$\beta$ ranges over $F_{p^n}$, $\Tr \beta$ takes each value in $F_p$ equally
often (i.e., $p^{n-1}$ times).
As we vary over $k$, $kx$ for $x \ne 0$ varies
over $F_{p^n}$ since $f(k): ~k \mapsto kx$ is a bijection.  So we can group the
sum into a sum of $p^{n-1}$ copies of $\sum_{\eta \in F_p} \omega^{\eta x}
= p \delta_{x,0}$, obtaining overall $p^n \delta_{x,0}$.  Thus   
(\ref{scooby dooby doo wah day})
becomes $\delta_{\alpha^2 - \beta^2 + \gamma^2 - \delta^2,0}~
\delta_{\alpha - \beta + \gamma - \delta,0}$.
So we have the simultaneous equations:
\beqa
{\alpha^2 - \beta^2 + \gamma^2 - \delta^2} = 0\;,
{\alpha - \beta + \gamma - \delta} = 0 \;
\eeqa
in $F_{p^n}$.
Rewriting these as
\beqa
(\alpha + \beta)(\alpha - \beta) = (\gamma + \delta)(\gamma - \delta) \\
(\alpha - \beta) = (\gamma - \delta)
\eeqa
we see that any $\alpha, \beta, \gamma, \delta$ satisfying
$\alpha = \beta, \gamma = \delta$ are solutions, and if one of
the latter conditions holds they both do.  If $\alpha \ne \beta$ (and so 
also $\gamma \ne \delta$), we can (since our arithmetic is in a field)
divide the first equation by 
the second to get the two equations $\alpha + \beta = \gamma + \delta$,
$\alpha - \beta = \gamma - \delta$, which are simultaneously satisfied
whenever $\alpha = \gamma, \beta = \delta$.  So if we write
\beqa
\Lambda = \sum \Lambda_{\alpha \gamma \beta \delta} 
\ket{\alpha}\bra{\beta} \otimes \ket{\gamma}\bra{\delta}
\eeqa
the matrix elements are 
\beqa \label{yabba dabba delta}
\delta_{\alpha \gamma} \delta_{\beta \delta}
+ \delta_{\alpha \beta}\delta_{\gamma \delta}\;,
\eeqa
except that each of the two terms in $(\ref{yabba dabba delta})$ gives
a unit
contribution when $\alpha = \gamma = \beta = \delta$, while the matrix
element $\Lambda_{\alpha \gamma \beta \delta}$
is still unity.  However, the full sum in (\ref{upsilondefined}),
including the standard
basis,  just adds an extra copy of precisely this
case, so that (up to normalization) 
$(\ref{yabba dabba delta})$ are the matrix elements of 
$\Upsilon$ in the standard basis.  These matrix elements are precisely 
those which define $\Pi$. \QED

I believe that the mutually unbiased bases 
defined by Wootters and Fields for $d=2^n$ also form spherical designs,
but have not shown it.  Indeed, it may be that any set of 
mutually unbiased bases necessarily forms a spherical 2-design.
The converse is true:  for a set of $d(d+1)$ vectors in $C^d$
to generate a spherical
2-design in $CP_{d-1}$, it is necessary that they be 
a set of $d+1$ MUBs.  (This follows from Theorem 44.9 in 
\cite{Hoggar96a}.)

These designs have an interesting relation to 
quantum error-correcting codes, and are also relevant in cryptography, 
where they serve to provide a {\em finite} ensemble with average-disturbance
properties similar to those for the uniform ensemble.  The information-disturbance
tradeoff is central to the power of quantum cryptography.  
The existence of such
finite ensembles may serve in some cases to allow specification of key or 
proto-key material with a finite amount of information, while retaining 
the strong
average-disturbance properties of states
completely unknown to one without the key information.
For example, these bases may serve to define the obvious $d(d+1)$-state 
generalization of the $6$-state protocol 
(\cite{Bruss97a}, \cite{Bruss98b})
on qubits.

\section{Conclusion}
We have defined and investigated properties of the information-disturbance
frontier for quantum measurements on an ensemble of states on a finite 
dimensional Hilbert space, as a particular way of formalizing the intuitive
notion that quantum mechanics often enforces a tradeoff between gaining
information and causing disturbance.  General properties of the frontier, such as
its convexity and monotonicity
were established.  

Specializing to important case of 
the uniform ensemble, representing a complete lack of 
knowledge about the initial state, we established further results
concerning information and disturbance.
For {\em any} measurement on this ensemble, we
showed that a least-disturbing way of doing it causes the 
system to suffer a dynamics, conditional on each measurement result,
described by a single Hellwig-Kraus operator.  We also established that
if we restrict ourselves to operations for which all Hellwig-Kraus operators
are positive (so that they represent the square-root conditional dynamics for
some measurement), a least-disturbing operation compatible with a given
measurement, for {\em any} ensemble, is to do the square-root dynamics
for that measurement:  fine-graining the measurement can never reduce
the disturbance.  However, we did not establish this for general conditional
dynamics, leaving as an interesting open question whether there are
non-uniform ensembles for which the least disturbing way of doing a particular
measurement is for the apparatus to collect additional information beyond
the measurement outcomes, and use it to aid in attempting to restore the 
initial state.  Our main result establishes that this is not so for the 
uniform ensemble, since an optimal instrument for measuring it just implements
the square-root conditional dynamics.  This allows us to calculate, for 
any measurement,  the minimal
disturbance compatible with a measuring it on the uniform ensemble.  This is 
only part of what is necessary to find the information-disturbance frontier,
which involves an in general difficult maximization of accessible information
subject to a disturbance constraint.  We showed that the maximal information
on the uniform ensemble may be obtained by unitarily covariant measurements 
and conditional dynamics.  Thus, the overall action of the measurement
dynamics on the state is just that of a 
``generalized depolarizing channel'' family of operations depending on
a single parameter $p$
which either do nothing to the state, or replace it with the maximally
mixed state with probability $p$.
It remains only to find the optimal measurement compatible with the
generalized depolarizing channel as a function of that parameter $p$.
Thus the  the problem of determining the information
disturbance frontier for the uniform ensemble is reduced from solving 
a parametric family of constrained maximization problems to solving 
a simpler parametric family of unconstrained ones. 

\section*{Acknowledgments}
Some of the work reported here was 
carried out at the Center for Advanced Studies of 
the University of New Mexico, and appeared in my UNM
doctoral dissertation \cite{Barnum98d}.
This work was supported in part by Office of Naval Research Grant
No.\ N00014-93-1-0116, National Science Foundation Grant
No.~PHY-9722614, the Institute for Scientific Interchange
Foundation, Turin, Italy and Elsag, a Finmeccanica company, 
and the 
European Union project QAIP, IST-1999-11234.
I thank Carlton M. Caves, Chris Fuchs, and Richard Jozsa for 
discussions, and an anonymous referee for valuable comments
and improvements.  The ideas herein originated from suggestions of
Chris Fuchs that one should look for an ``information-disturbance uncertainty
relation,''  and my interest in characterizing the projection postulate 
as somehow minimally disturbing, and were developed concurrently 
with Fuchs and Peres own development of the same formalism.
Richard Jozsa suggested the role of the
square root conditional dynamics.

\begin{appendix}
\section{Single-term conditional operations minimize uniform entanglement
disturbance} \label{appendix: min entanglement disturbance}
\begin{theorem}\label{theorem: single term operations 
and entanglement disturbance}
Let $\{F_b\}$ be a POVM and  
${\cal F}_b \sim \{ {F_b^{1/2}}\}$, and 
$\ca = \sum_b {\cal A}_b 
\sim \{ A_{bi} \}$, 
with 
$\sum_i A_{bi}^\dagger A_{bi}
 = F_b,$ 
be trace-preserving operations.
Then 
\beqa
F_e(I/d, \ca) \le F_e(I/d, {\cal F}_b)\;.
\eeqa 
\end{theorem}  

{\em Proof:}
We decompose ${\cal A}_b$ into the composition of two operations: a 
trace-decreasing operation ${\cal G}_b$ defined by:
\begin{eqnarray}
{\cal F}_b(\rho) = {F_b^{1/2}} \rho {F_b^{1/2}}
\end{eqnarray}
and an operation ${\cal B}_b$ 
(which is trace-preserving on the support
of $F_b$) defined by:
\begin{eqnarray}
{\cal B}_b(\rho) = \sum_i B_{bi} \rho B_{bi}^\dagger\;,
\end{eqnarray}
where $B_{bi} = A_{bi} F_b^{-1/2}.$  ($F_{bi}$ may not be invertible;
in this case, $F_{bi}^{-1/2}$ refers to the square root of the {\em generalized
inverse} of $F_{bi}.$  The generalized inverse is the inverse on $F_{bi}$'s 
support (where it is invertible)
extended (as a direct sum) by the zero operator on the orthocomplement of
the support.)  It is easily seen that
${\cal B}_b$ is trace-preserving on 
$F_b$'s support  ($\sum_i B_i^\dagger B_i = \Pi_b$, where
$\Pi_b$ is the projector onto the support of $F_b$),
and that ${\cal B}_b \circ {\cal F}_b = {\cal A}_b.$  

Then 
\begin{eqnarray}
F_e(I/d, {\cal A}) = \frac{1}{d^2}\sum_{bi} |{\rm tr} B_{bi} F_b^{1/2}|^2.
\end{eqnarray}
By the Schwarz inequality,
\begin{eqnarray}
|{\rm tr} B_{bi} F_b^{1/2}|^2 \equiv |{\rm tr} B_{bi} F_b^{1/4} F_b^{1/4}|^2 
\le ({\rm tr} B_{bi} F_b^{1/2} B_{bi}^\dagger)( {\rm tr} F_b^{1/2})\;, 
\end{eqnarray}
so by the trace-preserving property for each ${\cal B}_b,$
\begin{eqnarray}
F_e(I/d, \ca) \le \frac{1}{d^2}\sum_b |{\rm tr} F_b^{1/2}|^2 = F_e(I/d, {\cal G}).
\end{eqnarray}
This is just the entanglement fidelity for the uniform density operator
when the operation ${\cal G}$ corresponding to the generalized L\"uders'
rule is used.  Hence the generalized projection postulate minimizes
disturbance to the entanglement of the uniform density operator.

\section{More on one-term versus multi-term conditional operations}
\label{appendix: towards the possibility}
Here I consider some other approaches towards proving Theorem 
\ref{oop bop sh'bam}.  These have so far proven unsuccessful
except in the case in which all POVM elements commute.
They are still of some  interest in that they attempt to establish
intermediate results stronger than Proposition \ref{oh frabjous day}.

In the one-term conditional dynamics
case, we had:
\beqa
|\melement{\psi}{A_b}|^2 \le | \melement{\psi}{F_b} |^2 \;.
\eeqa
In the multiple-term conditional
dynamics case, we might hope to establish that
\beqa
\label{oompa-loompa}
\sum_{i}  |\melement{\psi}{A_{bi}}|^2 \le |\melement{\psi}{F_b}|^2\;.
\eeqa
If $A_{bi}$ is assumed positive (as it is for the conditional dynamics
which are minimally-disturbing to the uniform ensemble), this follows from Proposition 
\ref{oh frabjous day}, but we might try to establish \ref{oompa-loompa}
without that assumption.  (There is no hope of estabilishing that $U_{bi} = I$, 
i.e. $A_{bi}$ positive, is minimally disturbing for an arbitary ensemble;  it 
is obviously not true, for example, when the ensemble has all probability concentrated
on one state $\ket{\psi}$.)

Defining $\cb_b$ and $B_{i}$ as in Appendix 
\ref{appendix: min entanglement disturbance},
\begin{eqnarray}
|\langle \psi | A_{bi} |\psi \rangle|^2 &=& 
|\langle \psi | B_{bi} F_b^{1/2} |\psi \rangle|^2 \nonumber \\ 
&=&|\langle \psi | B_{bi} F_b^{1/4} F_b^{1/4}|\psi\rangle|^2\;.\label{eq: preschwarz}
\end{eqnarray}
Applying the Schwarz inequality as before gives
\beqa
&&|\langle \psi | B_{bi} F_b^{1/4} F_b^{1/4}|\psi\rangle|^2\; \nonumber \\
&\le& \; \langle \psi | B_{bi} F_{b}^{1/2}B_{bi}^\dagger |\psi\rangle
\langle \psi | F_{b}^{1/2}|\psi\rangle\;. \label{eq: postschwarz}
\eeqa
If the inner product were a trace, as before, we would just 
cycle $B_{bi}^\dagger$ next to $B_{bi}$ and then sum on 
$i$ to get the identity, removing the $B$'s entirely and establishing
equation (\ref{oompa-loompa}).  Unfortunately,
we cannot do that here unless $B_{bi}$ commutes with $F_b$.  
Nor is it clear we can cycle one
of the 
$F_b^{1/4}$ in (\ref{eq: preschwarz}) around to give 
$|\langle \psi | F_b^{1/4}B_{bi} F_b^{1/4} |\psi\rangle|^2\;,$
which would have given rise to the desired ordering after the
Schwarz inequality was applied.  To proceed from 
\ref{eq: postschwarz} means we are trying to show that:
\beqa
\sum_{i}\langle \psi | B_{bi} F_{b}^{1/2}B_{bi}^\dagger |\psi\rangle
\le 
\langle \psi | F_{b}^{1/2}|\psi\rangle\;,
\eeqa
using the fact that $\sum_{i} B_{bi}^\dagger B_{bi} = I$.  
However, counterexamples to 
\beqa \label{eq: hopeless}
|\langle \psi | {\cal E} (G) |\psi \rangle|
\le 
|\langle \psi | G |\psi \rangle|
\eeqa
for trace-preserving ${\cal E}$ and $0 \le G \le I$
are easily found.  For example, let $G$ be proportional to
a projector onto some
state other than $|\psi\rangle$, and
let ${\cal E}$ unitarily rotate that state back to $|\psi\rangle$.
Since $\cb_b$ may be an arbitrary trace-preserving operation, 
this means that the Schwarz inequality as applied to obtain
equation (\ref{eq: postschwarz}) is too loose for our purposes, and
we must work with (\ref{eq: preschwarz}), summed over $i$.

In fact, the counterexample to (\ref{eq: hopeless}) given above also
shows that even this will not work:  there is no hope of establishing
equation (\ref{oompa-loompa}), because it is equivalent to:
\beqa
\melement{\psi}{\cb \cf (\proj{\psi})} \le 
\melement{\psi}{\cf (\proj{\psi})} \;.
\eeqa
Rather, we might try to show that
\beqa
\melement{\psi}{\cb \cf (\proj{\psi})} \le
\max_{\rm unitary~U} \melement{\psi}{U \cf (\proj{\psi}) U^\dagger} \;,
\eeqa
\end{appendix}
where $\cb$ is arbitrary and trace-preserving and
$\cf \sim \{F\}, I \ge F \ge 0$.


\end{document}